\documentclass[peerreview,10pt,letterpaper]{IEEEtran} 

\usepackage[utf8]{inputenc} 

\usepackage{geometry} 

\usepackage{graphicx} 
\usepackage{xcolor}

\newcommand{\hydra}{\textit{Hydra}}
\newcommand{\cretin}{\textit{Cretin}}
\newcommand{\rsquare}{$\text{R}^2$}
\newcommand{\change}[1]{\textcolor{black}{#1}}
\newcommand{\changeII}[1]{\textcolor{black}{#1}}

\usepackage{booktabs} 
\usepackage{array} 
\usepackage{paralist} 
\usepackage{verbatim} 
\usepackage{float}
\usepackage{amsmath}
\usepackage{multirow}

\begin{document}
\title{Transfer Learning of High-Fidelity Opacity Spectra in Autoencoders and Surrogate Models}
\author{Michael~D.~Vander~Wal,
        Ryan~G.~McClarren,
        Kelli~D.~Humbird,
\thanks{M. D. Vander Wal and R. G. McClarren are with the Aerospace and Mechanical Engineering Department, University of Notre Dame, Notre Dame, IN, 46556 USA, email: Michael~D.~Vander~Wal - mvander5@nd.edu; Ryan~G.~McClarren - rmcclarr@nd.edu.}%
\thanks{K. D. Humbird is with Lawrence Livermore National Laboratory, 7000 East Ave, Livermore, CA, 94550 USA, email: humbird1@llnl.gov.}}%

\maketitle

\begin{abstract}
Simulations of high energy density physics are expensive, largely in part for the need to produce non-local thermodynamic equilibrium opacities. High-fidelity spectra may reveal new physics in the simulations not seen with low-fidelity spectra, but the cost of these simulations also scale with the level of fidelity of the opacities being used. Neural networks are capable of reproducing these spectra, but neural networks need data to to train them which limits the level of fidelity of the training data. This paper demonstrates that it is possible to reproduce high-fidelity spectra with median errors in the realm of 3\% to 4\% using as few as 50 samples of high-fidelity Krypton data by performing transfer learning on a neural network trained on many times more low-fidelity data.
\end{abstract}

\begin{IEEEkeywords}
multi-fidelity, opacities, transfer learning, djinn, NLTE
\end{IEEEkeywords}

\IEEEpeerreviewmaketitle

\section{Introduction}
Inertial confinement fusion (ICF) is currently one of the experimental approaches to controlled nuclear fusion. ICF involves the heating and compressing of a  deuterium and tritium (DT) fuel target. This is generally accomplished with lasers which are used to rapidly heat the fuel either directly or indirectly such that shock waves are formed in the fuel which compress and heat the fuel further. Indicated by the name of the process, lasers directly illuminate the fuel in direct-drive ICF. In the case of indirect-drive ICF, lasers heat the inside of a hohlraum which subsequently causes the generation of the high-intensity x-rays. The x-rays then heat and compress the fuel capsule, which sits inside of the hohlraum \cite{kline,li2011,dodd2018}.

ICF experiments are exquisitely complex to field and expensive to execute. Each ``shot'', as the experiments are often called, can easily cost in the realms \$1 million. This means, of course, that experiments are not performed lightly, and significant effort goes into designing the experiments using computational models.


Computer simulations are comparatively cheap, and are thus useful tools for scoping out vast design spaces to search for optimal experimental settings. Simulations of ICF implosions are multiphysics simulations, involving hydrodynamics, radiation transport, nuclear fusion burn, equations of state at extreme conditions, and more. Depending on the degree of accuracy needed in each physics subpackage, approximations can be made or tabled data may be used to reduce the complexity of an integrated simulation. For example, opacities can be tabulated for assumptions of local thermodynamic equilibrium, but there are regimes in ICF experiments that violate these assumptions. In this case, higher fidelity opacity calculations are necessary to capture important physical processes accurately \cite{holladay2020accelerated,kluth}. In this work, we focus on improving the fidelity of opacity calculations without significantly increasing the computational cost of the overall ICF simulation using machine learning. 

The two primary components of the opacity are the absorptivity and emissivity. At conditions one might generally encounter throughout everyday life, these properties are considered to be equal and only dependent on temperature \changeII{and density} as prescribed by the blackbody assumption and fall under the realm of local thermal equilibrium (LTE) calculations \cite{holladay2020accelerated,kluth,dodd2018}. ICF conditions, however, demand that non-local thermal equilibrium (NLTE) calculations be made. NLTE calculations are dependent on both the density and temperature of the material in question \change{as well as the radiation- and electron-energy distributions. Unlike LTE conditions where the relevant radiation interaction parameters can be tabulated and referenced during a calculation, such tables for NLTE cannot be easily constructed, except for in special, limited circumstances \cite{scott2005non,fontes2000non} or using simplified models \cite{takabe1994computational,hu2012mitigating}. Recent work \cite{walton2020parameterizing} has developed a parameterization scheme for the electron distributions to simulate NLTE K$\alpha$ and K$\alpha$ emission spectroscopy. This is a potentially fruitful area for further development, but at present has not been demonstrated for the laser-hohlraum interactions that we consider here.} 

\begin{figure}[!t]
    \centering
    \includegraphics[width=\linewidth]{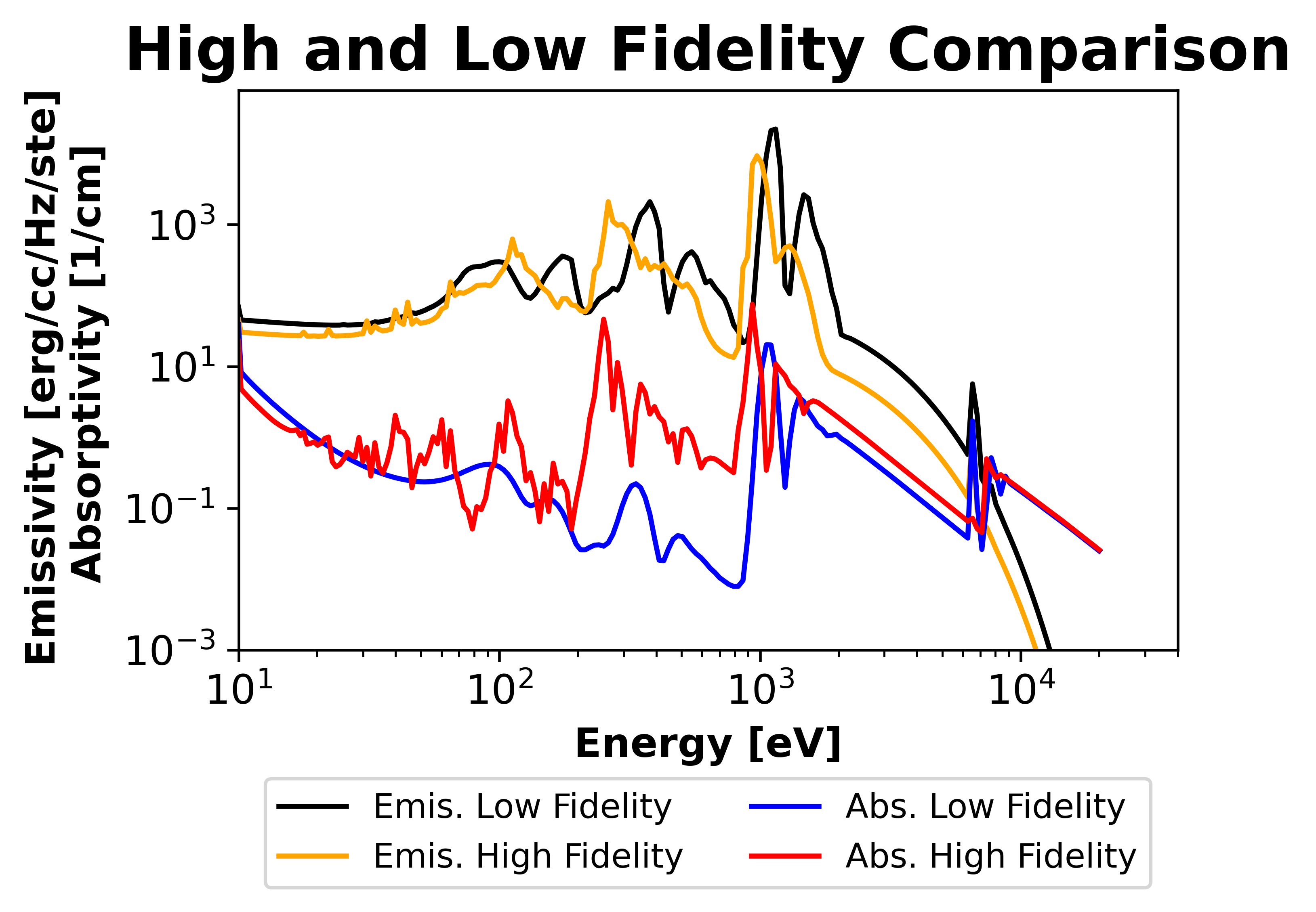}
    \caption{\change{An example of the differences between  the spectra computed using low- or high-fidelity atomic models in \cretin{} for a Kr plasma. For the emissivity curves, the density is $.003083$ g/cm$^3$, an electron temperature of $2045.3067$ eV, a radiation temperature of $288.242734$ eV, and an M-band ratio of $\alpha = .236522$. The absorptivity shown corresponds to a density of $.003145$ g/cm$^3$, an electron temperature of $314.909732$ eV, a radiation temperature of $299.114323$ eV, and an M-band ratio of $\alpha = .170141.$}}
    \label{fig:lf compare}
\end{figure}

NLTE calculations are significantly more complex than LTE calculations, and thus have a much higher computational cost. Indeed, this cost is so high that the calculation of the spectral absorptivity and spectral emissivity may constitute as much as 90\% of the total computation time of an ICF hohlraum simulation. This cost is dictated by the level of fidelity of the model where the fineness of energy-binning as well how many electronic states will be used to determine the level of fidelity. The difference in these levels manifests such that low fidelity models end up with many spectral lines or spikes being grouped together and effectively being smoothed out as seen in Figure \ref{fig:lf compare}. This is particularly apparent in low temperature conditions as well as low density conditions. The former is effectively the initial condition of the simulation. In some cases, the level of fidelity desired may require a whole day to complete a single calculation \cite{kluth}. This level of fidelity is not currently used to run full ICF simulations. The fidelity of the models used in the actual ICF simulations is necessarily much lower; however, they are still expensive.

This is where machine learning, or more specifically neural networks, become exceptionally useful. Neural networks can be considered universal function approximators \cite{scarselli1998,cybenko1989,hornik1988}. Neural networks, once trained, provide an efficient route to calculate any desired function or set of functions. This is because neural networks can provide an answer without the iterative solving of equations. Additionally, where some solvers are not well suited for parallelization, neural networks are embarrassingly parallel \cite{raghavendra2007}.

Indeed, it has been demonstrated that a 7x speed-up was observed for simulations in \hydra{} where a set of neural networks replaced \cretin{} \cite{kluth}.  \hydra{} is the hydrodynamics multi-physics simulation software most commonly used for ICF simulations at Lawrence Livermore National Laboratory \cite{hydra}. \cretin{} is the standard package used by \hydra{} to calculate the NLTE opacities of materials \cite{cretin}. It was also demonstrated that multiple elements could be included in a single model \cite{vanderwal1}. \changeII{An important finding from that work is that data needs to be properly scaled to allow neural networks working in single precision arithmetic to capture the large range of atomic spectra.} Both of these examples utilized large datasets based on low-fidelity NLTE calculations, and do not address the desire to use the time-cost prohibitive high-fidelity models. If this high-fidelity data could be calculated with a neural network, the 7x speed-up could very well become a 75x speed-up or higher relative to the direct computation. This expectation considers that the low-fidelity models used in the proof-of-concept take less than one second to compute a single set of opacity calculations. The challenge with applying this technique to high fidelity opacity data is that acquiring the same amount of data required for  training the low-fidelity network is itself prohibitively expensive.

Transfer learning is a proposed method to reduce the computational time required to train an accurate high fidelity neural network. Transfer learning is any of various techniques that take an already trained neural network and reuses all or only parts of the neural network to achieve either similar or potentially very different tasks \cite{mcclarren2021machine}. Transfer learning is also used to compensate for small datasets \cite{zhuang2021,meng2020,aydin2019,luo2018,zhu2019,zhuang2015,deng2013,deng2014,eusebio2018,imai2020}. 

This paper explores the use of various transfer learning techniques to achieve the goal of predicting high-fidelity opacities with similar accuracy as a low-fidelity opacity model while using a comparatively small dataset. The methods that are explored include simple retraining on the new data, retraining on the new data with various collections of layers being frozen, and linking a pre-trained low-fidelity network to an untrained high-fidelity network \cite{meng2020}.  The ultimate purpose of this work is to demonstrate that transfer learning can be used to effectively learn high-fidelity opacity models, as well as give direction and suggestions on how to perform this task.

\subsection{Prior Work and Contributions}
Transfer learning with autoencoders has been performed for a wide variety of tasks. These often include the use of an autoencoder to learn existing structures. The autoencoder's decoder is then replaced with a new network that performs a desired task such as classification \cite{luo2018,zhu2019,zhuang2015,deng2013,deng2014,eusebio2018}. In a few cases, there has been work that has not been directly attributed as being transfer learning but does closely resemble transfer learning as a pre-exisiting network is used \cite{meng2020,aydin2019}. In a manner similar to the use of transfer learning from autoencoders to produce classification networks, the work in \cite{kluth,vanderwal1} replace the classification networks with regression networks. Once again, they are not directly attributed to the process of transfer learning, but are similar in nature. In general, there appears to be less in the way of transfer learning for the purposes of regression \cite{meng2020,aydin2019,kustowski2019,goswami2020,humbird2021,humbird2020}. Indeed, some of those cases were not even directly attributed as transfer learning are among the small amount of work done in this area. The work shown in \cite{aydin2019} describes the approach taken in this paper as a ``brute force" approach. 

Utilizing neural networks to predict spectra or material properties is not itself new either. There has been much work on predicting the spectra of chemical or elemental species as well as the inverse problem: the prediction of which chemical or elemental species are present based on observed spectra \cite{ghosh2019,chatzidakis2019,liu2019,cui2018}. Also in the field of chemistry, transfer learning has been used to improve to the performance of autoencoders in generating new chemicals \cite{iovanac2020}. 

The work in this paper is an extension of efforts to reproduce spectra -- absorptivity and emissivity -- for non-local thermal equilibrium conditions. In our initial work, autoencoders are trained to compress and decompress the spectra, and then Deep Jointly Informed Neural Network (DJINN) models are trained to predict the latent spaces of those autoencoders. During prediction, the DJINN model takes the inputs to produce the latent space representation that the autoencoder's encoder would have produced. The autoencoder's decoder then decompresses the predicted latent space into the predicted spectra \cite{kluth,djinn}.
The contributions of this paper are the demonstration that high-fidelity models can be learned by the process of transfer learning from low-fidelity models that a median relative errors on the order of 3\% to 4\% can be achieved, and that this can be obtained with as few as 50 high-fidelity samples.


\section{Methods}
\subsection{Data}
The training data is generated using the code \cretin{}. \change{For both  the low-fidelity and high-fidelity calculations, \cretin{} solves the equations of the collisional-radiative model in steady-state using a screened hydrogenic model \cite{scott2010advances}. The difference between the calculations is that the low-fidelity calculations use 1\,849 possible electronic configurations of the krypton atom whereas the high-fidelity calculations use 25\,903 levels following \cite{scott2010advances}.} The inputs into \cretin{} are the density ($\rho$), electron temperature ($T_e$), and the radiative field. The radiative field provided to \cretin{} is generated by producing a \change{superposition of a Plank distribution  with a radiative temperature ($T_r$) and a Gaussian representing M-band radiation}:
\begin{equation}
\label{eq:rad field}
I(\nu) = aT^{4}_r \left[\left( 1-\alpha\right)b\left(\nu,T_r\right) \, + \, \alpha g\left(\nu \right) \right].    
\end{equation}
In this expression, $a$ is the radiative constant of $7.5657\times 10^{-15}$ $\text{erg}/\text{cm}^3/\text{K}^4$, $b(\nu,T_r)$ is the reduced Planckian, and $g(\nu)$ is a Gaussian distribution centered at 3 keV with a full width half maximum set to 1 keV, and $\alpha$ is the M-band ratio \change{that gives the fraction of the radiation field in the M-band between 2-4 keV characteristic of gold} \cite{kluth,li2011}. \change{The spectral outputs from \cretin{} (emissivity and absorptivity) are stored in  200 bins ranging up to 40 keV and arranged to capture the K and L edges.}  All inputs are sampled from uniform distributions: $\rho$ is sampled from 0.003 g/cm\textsuperscript{3} to 0.1 g/cm\textsuperscript{3}, $T_e$ is sampled from 300 eV to 3000 eV, $T_r$ is sampled from between 30 eV and 300 eV, and $\alpha$ is sampled from 0.0 to 0.3.

Using these randomly generated inputs, 320,000 low-fidelity krypton spectra training samples are generated each for absorptivity and emissivity. For training the neural networks, each separate model has its own .8/.2 train/test split of the 320,000 samples. The 20\% used for testing will never be used to train the model that is associated with that specific split.

The high-fidelity training spectra are produced with a more detailed atomic model in \cretin, with sample sizes of 25, 50, 100, 250, 500, 1000, and 2000. The high-fidelity data is sampled using Latin hypercube sampling (LHS) with the same parameter boundaries defined above. \change{LHS is used to sample the parameter space to ensure that the parameter space range is well-sampled when the number of samples is small \cite{santner2003design,mcclarrenUQ}.} To give perspective on the difference in computation cost between the low-fidelity and high-fidelity models, the low-fidelity model generates approximately 33 spectra per minute, and the high-fidelity model generates approximately 3 spectra per minute. An additional set of 60,000 test spectra are generated using the high-fidelity model with  random sampling.


\subsection{Basic Retraining}
\change{Our work is based on deep neural networks designed to represent the complex atomic physics calculations. Neural networks are comprised of a set of intermediate calculations, usually referred to as neurons, that take input data and perform a calculation to produce an output. The connections between the neurons (and how the neurons combine data) are determine through a set of free parameters, the so-called weights and biases of the network, that are determined through a training process where these parameters are set in order to match existing data known as training data. Additionally, there are several hyperparameters that the network designer must choose. These include the number and connectivity of the neurons that give the network architecture as well as activation functions for the neurons. One such architecture that we use heavily is the autoencoder that seeks to find a low-dimensional subspace on which a high-dimensional dataset lives. Additionally, the aforementioned DJINN method attempts to automatically select many of the hyperparameters for a given network based on the training data. The methods we use in this work have been well-described in several recent publications \cite{humbird2020,kluth,vanderwal1, mcclarren2021machine} and the interested reader is encourage to consult these for additional background. }

First, two different autoencoder architectures are trained on low-fidelity data. One architecture is fully-connected and is described in Table \ref{tab:mlp ae}, and the other architecture is convolutional, described in Table \ref{tab:conv ae}. These autoencoders are trained to compress and decompress the absorptivity and emissivity spectra with separate autoencoders used for each. Separate DJINN models are trained to reproduce the low-fidelity autoencoders' latent spaces. In total there are ten complete models for absorptivity and emissivity for each variation of model used.

\begin{table}[]
\centering
\caption{The fully connected autoencoder architecture.}
\label{tab:mlp ae}
\begin{tabular}{rc}
\toprule
Input Dim  & 200        \\
Encoder    & 100,33     \\
Latent Dim & 10         \\
Decoder    & 33,100,200 \\
Activation & softplus  \\
\bottomrule
\end{tabular}
\end{table}

\begin{table}[]
\centering
\caption{The convolutional autoencoder architecture.}
\label{tab:conv ae}
\begin{tabular}{rc}
\toprule
Input Dim      & 200                                                                          \\
Encoder Conv   & \begin{tabular}[c]{@{}c@{}}30 filters, size = 1x5,\\ stride = 1\end{tabular} \\
Encoder        & 75, 50, 30                                                                   \\
Latent Dim     & 15                                                                           \\
Decoder        & 30, 50, 275, 6000                                                            \\
Decoder Deconv & \begin{tabular}[c]{@{}c@{}}1 filter, size = 1x5,\\ stride = 1\end{tabular}   \\
Output Dim     & 200                                                                          \\
Activation     & \begin{tabular}[c]{@{}c@{}}exponential linear\\ unit\end{tabular} \\
\bottomrule
\end{tabular}
\end{table}

The low-fidelity autoencoders are trained on a .9/.1 split during the actual training session. The 10\% holdout is used to produce validation cost. This means that each low-fidelity model is trained on 230,400 samples. Prior to training, the spectra are scaled by taking the 18th-root of all bins in the spectra. The fully-connected autoencoders are trained for 10,000 epochs with a learning rate of 0.001 while using mean squared error (MSE) as the loss function. The batch size is 1,280 spectra per batch. All of these training details are the same for the convolutional autoencoders with the exception of the learning rate, which is 0.0001.

The DJINN models are then trained to reproduce the latent space of the autoencoders given the inputs associated with the corresponding spectra. DJINN automatically constructs a model using a few hyper-parameters. Those parameters are dropout rate, maximum number of layers, and the number of models to use in an ensemble. The dropout rate was set to 0.0, the maximum number of layers to 11, and the number of models to 1. By the very nature of DJINN, the actual number of layers is not guaranteed to be equal to the maximum; however, in this case, all models reach the maximum number of layers. DJINN employs MinMax scaling for its inputs and outputs, and it produces models that have layers that all use relu activation functions with the exception of the last layer which uses a linear activation function. DJINN also uses MSE as the loss function. The DJINN models are trained for 2,000 epochs with the learning rate set to 0.0001, and the batch size of 1,280 spectra per batch.

The autoencoders are trained on the high-fidelity spectra with various alterations made to the autoencoders. These alterations consisted of freezing different layers of the autoencoders, such that they could not be retrained. For clarity, we define each model by the layers which are trainable during the transfer learning step, as it is more compact to state those than the layers that are frozen. Table \ref{tab:freeze layers} summarizes the trainable layer combinations as well as the labels attributed to each of them. There is also an instance of no layers being frozen as well as a brand new (randomly initialized) model produced on just the high-fidelity spectra.

\begin{table}[]
\centering
\caption{The labels associate with the layers left not frozen.}
\label{tab:freeze layers}
\begin{tabular}{rc}
\toprule
Label             & Trainable Layers                                                                       \\
\midrule
NO FREEZE         & all layers                                                                             \\
FLATL             & first, latent, last                                                                    \\
FIRSTLAST         & first, last                                                                            \\
LATENT            & latent                                                                                 \\
LAST              & last                                                                                   \\
LATLAST           & latent, last                                                                           \\
NO FREEZE - CONV  & all layers                                                                             \\
NO CONVALT - CONV & \begin{tabular}[c]{@{}c@{}}first, second, latent, \\ second to last, last\end{tabular} \\
\bottomrule
\end{tabular}
\end{table}

The hyperparameters used for transfer learning differ from those used to train the low fidelity models. The 25, 50, and 100 spectra datasets use batch sizes of 1. The 250, 500, 1,000, and 2,000 spectra dataset use 2-, 3-, 5-, and 10-spectra batch sizes. There wisas also no .9/.1 validation split. The models are trained for 2,000 epochs on the high-fidelity data using MSE as the loss function. The learning rates of both the fully-connected and convolutional models are reduced by a factor of ten, making them 0.0001 and 0.00001, respectively.

Each of the autoencoders produced via the different combinations of frozen-unfrozen layers had DJINN models to match that are either new, continued training of the unaltered model, had the last two layers trainable, or had just the last layer trainable. Ten models are produced for each combination of the autoencoders and DJINN models. The DJINN models produced here are trained with the same batch sizes as their associated autoencoders. The models are trained for 1,000 epochs on the high-fidelity latent spaces using a learning rate that is similarly a factor of ten smaller at 0.00001. It should be noted that the new DJINN models during this step are also built with a maximum number of layers set to 11, but, because the number of layers actually produced is related to amount of training samples available at construction time, the actual number of layers tends to be much less than 11.

A very important note for the success in the implementation of neural network models is the scaling of the input and output data of the neural network. The data covers a very large range of values, especially the emissivity spectra, in which values drop lower than $10^{-40}$, and the neural networks have an apparent floor around $10^{-4}$. \changeII{As discussed in detail in our previous work, this large range requires a compression of the order of magnitude of the data \cite{vanderwal1}. The data here are compressed by taking the eighteenth-root of the data. Compression of the range is necessary because our neural networks use single precision arithmetic. This is done for the purpose of speed due to the fact that most hardware for ML is tuned for single precision. In single precision models, our previous work showed neural networks could not reproduce results that were more than 7 orders of magnitude smaller that the mean of the data \cite{vanderwal1}.  }

\section{Results}

Before detailed explanation of the results a baseline must be set. The baseline for comparison is the low-fidelity model's ability to reproduce the high-fidelity results. If the transfer-learned models cannot reproduce the high-fidelity results better than the low-fidelity then there is no point in trying. \changeII{This is not an unreasonable question. The low fidelity atomic models were created to minimize the error in the results when going from a large number of levels in the high-fidelity model to a smaller model. Therefore, in many regions of parameter space, the low-fidelity and high-fidelity results may be very similar.  It is important to remember, however, that even the low-fidelity atomic physics calculations are much slower than evaluating a trained neural network \cite{kluth}.} 

\changeII{The results from comparing transfer learning to the low-fidelity model} can be seen in Table \ref{tab:low fidelity results}. The errors reported throughout this paper are all percent relative error as well as the \rsquare value. The errors are presented as a spectral error, meaning that the value represents one specific energy bin in a specific spectra. This is different from how error was reported in \cite{kluth,vanderwal1} where it was reported on a gray approximation of the spectra. The \rsquare{} value is used as a measure to indicate how well the shapes of the spectra are reproduced. Unless otherwise specified, the metrics reported: median, 90th percentile, and maximum, are all the median values of those metrics across the ten models produced for a given combination of autoencoder and DJINN model. This is the same for the \rsquare{} value with exception that 90th-percentile will be considered from an inverted ordering (i.e. 90\% have a greater \rsquare{} value as opposed to less value).

Accompanying the low-fidelity results in Table \ref{tab:low fidelity results} are the results of a selected fully-connected model combination and a convolutional model combination. They are also accompanied by the actual low-fidelity data associated with the same inputs as the high-fidelity data. The fully-connected model is the model made with a ``NO FREEZE" fully-connected autoencoder trained on 50 samples, and the convolutional model is the model made with a ``NO FREEZE" convolutional autoencoder trained on 50 samples as well. Both of the models utilized a ``NO FREEZE" DJINN model. The convolutional model performs better than the fully-connected model for both the median and 90th percentile errors, achieving 2.55\% and 3.81\% median errors, for both absorptivity and emissivity respectively. 

The \rsquare{} values provide a further point of differentiation between the fully-connected model and the convolutional models. The convolutional models perform better than the fully-connected, absorptivity models, but if the median and 90th-percentile errors are ignored, the convolutional models can be realized to perform worse for emissivity. This is especially apparent if the maximum relative error is of particular importance. 

There is the unfortunate reality that the minimum \rsquare{} value for the absorptivity models is negative, but the one calculated is still a better value than that of the low-fidelity data. The \rsquare{} value for the emissivity models may give rise to some concern; however, the primary reason for this is because the values for the energy bins greater than roughly 8 keV have an even more rapid drop off in value for the high-fidelity data than the low-fidelity data. This is just above the range of energy values of primary importance. The emissivity models otherwise have very good shape matching capacity.

Ultimately, these results also show that both the fully-connected models and convolutional models outperform their low-fidelity model counterparts in reproducing high-fidelity data. The high-fidelity models roughly halve both the median and the 90th-percentile relative errors. The low-fidelity data already has a relatively good shape match and in some cases better errors, but it tends to be at the extrema of spectra where the bulk majority of this increased error appears. Additionally, using more samples does not appear to improve errors or shape match much. Using 2,000 samples, the 90th-percentile \rsquare{} value was 0.9954 for the fully-connected emissivity model, and the 90th-percentile \rsquare{} of the convolutional absorptivity model was 0.9534. Neither of these values suggest that using more samples will provide meaningful benefit.

\begin{table*}[!t]
    \centering
    \caption{The low-fidelity data, low-fidelity model results, and transfer learned model results relative to the high-fidelity data.}
    \label{tab:low fidelity results}
    \begin{tabular}{rccccc}
    \toprule
    Absorptivity    & LF-DATA  & LF-FC     & LF-CONV   & TL-FC     & TL-CONV   \\
    \midrule
    Median               & 2.46        & 7.85        & 76.84         & 3.20          & 2.55          \\
90th Percentile      & 22.3        & 41.9        & 40.0          & 20.3          & 19.9          \\
Maximum              & $1.80\times 10^4$ & $2.10\times 10^4$ & $1.84\times 10^{4}$ & $1.94\times 10^{4}$ & $1.83\times 10^{4}$ \\
\rsquare{} Median          & 0.9971      & 0.9826           & 0.9826             & 0.9988        & 0.9990        \\
\rsquare{} 90th Percentile & 0.9536      & 0.9028           & 0.9031             & 0.9526        & 0.9513        \\
\rsquare{} Minimum         & -0.3733     & -0.2678           & -0.2100             & -0.2547       & -0.1784 \\
    \bottomrule
    \end{tabular}
    \begin{tabular}{rccccc}
    \toprule
    Emissivity      & LF-DATA     & LF-FC       & LF-CONV        & TL-FC          & TL-CONV        \\
    \midrule
Median               & 3.19        & 8.04        & 7.87           & 4.32           & 3.81           \\
90th Percentile      & 32.8        & 70.9        & 72.7           & 37.0           & 36.9           \\
Maximum              & $5.04\times 10^5$ & $4.44\times 10^8$ & $1.95\times 10^{18}$ & $2.26\times 10^{11}$ & $7.82\times 10^{20}$ \\
\rsquare{} Median          & 0.9999       & .9979           & .9978              & 0.9998          & 0.9998          \\
\rsquare{} 90th Percentile & 0.9961       & .9909           & .9905              & 0.9951          & 0.9944          \\
\rsquare{} Minimum         & 0.8921       & .8778           & .8034              & 0.8822          & 0.7619    \\
    \bottomrule
\end{tabular}
\end{table*}

\begin{figure}[!t]
    \centering
        \includegraphics[width=\linewidth]{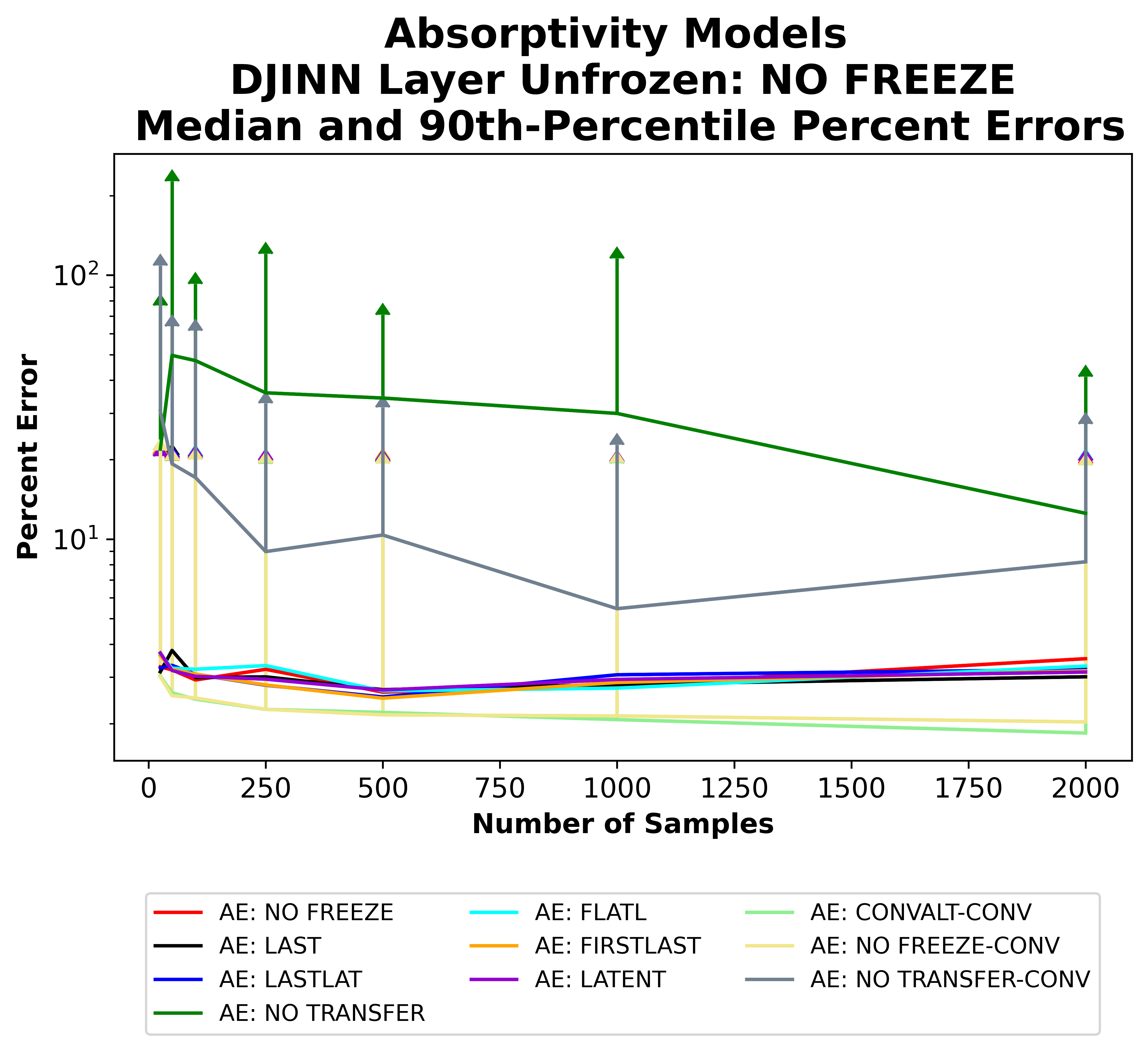}
        \includegraphics[width=\linewidth]{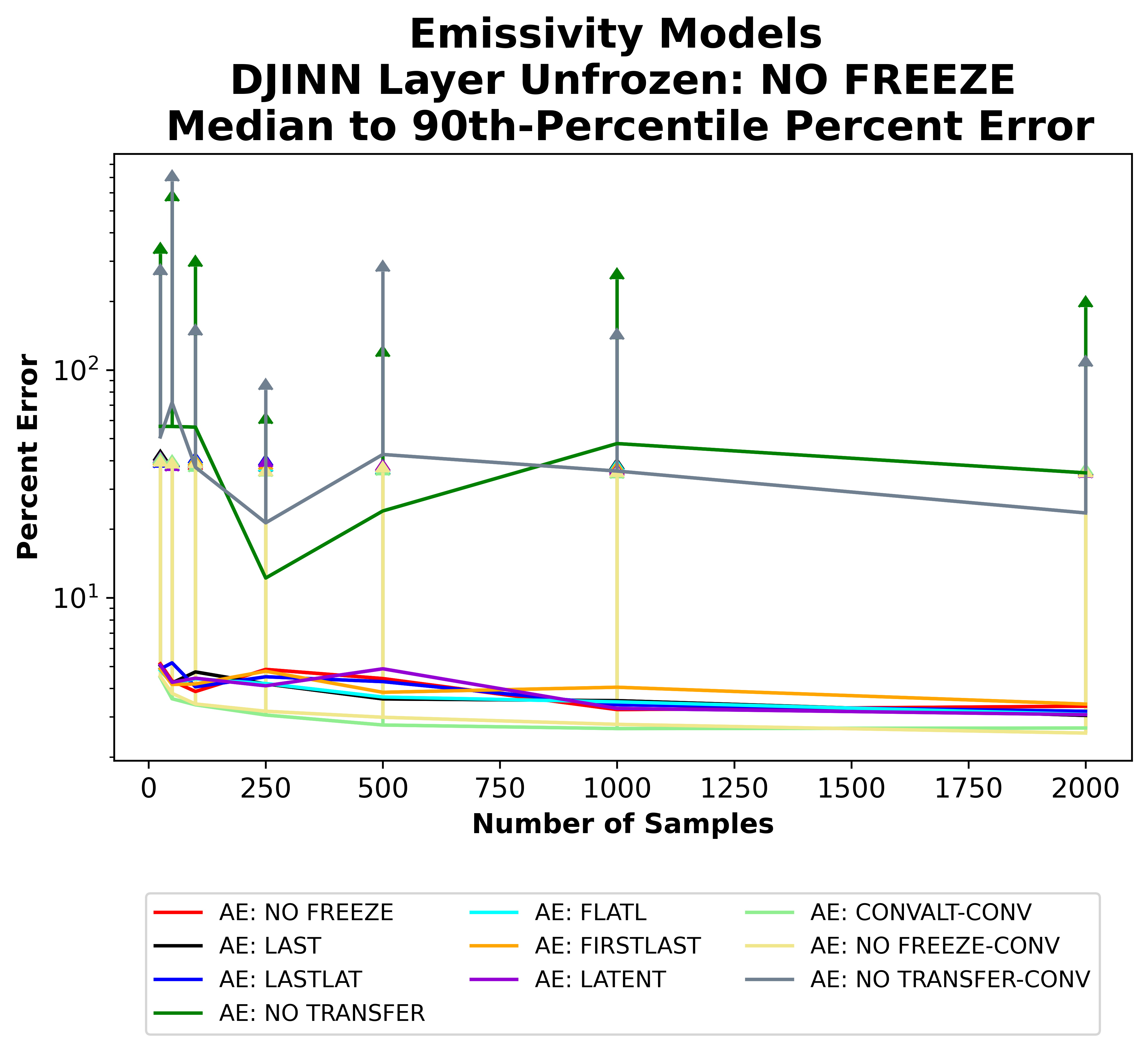}
    \caption{The results of models made a ``NO FREEZE" DJINN model and the various different autoencoder options. The tight clustering of the lines in these plots make it clear that the freezing of autoencoder layers does not play a clear or major role. However, the bottom most lines show that convolutional autoencoders perform better. The vertical lines is the 90th-percentile error. The error values reported are the median values of each reported metric (median or 90th-percentile) out of ten separate models.}
    \label{fig:dj no freeze median}
\end{figure}

The bottom most lines in the plots in Figure \ref{fig:dj no freeze median} further show how the convolutional models produce the lowest median errors, and if provided with increasing amounts of data the median error continues to drop. This is in contrast to the fully-connected models, represented by the tight clustering of lines in the plot, which increase in median error when more than 500 samples are provided. Ultimately, these lines show that the choice of trainable layers does not particularly matter for a particular class of autoencoder, fully-connected versus convolutional. The same goes for emissivity provided that there are enough samples. The top two lines of the plots in \ref{fig:dj no freeze median} are from the models with a transfer learned DJINN model but a brand new autoencoder. The significantly worse performance for these two models can be explained by the fact that the DJINN model was originally trained on a different latent space because the autoencoder is different.


 One last interesting thing to note is that for absorptivity and, to a more limited extent, emissivity, is that using more samples does not guarantee better median errors or even maximum errors. 

\begin{figure}[!t]
    \centering
        \includegraphics[width=\linewidth]{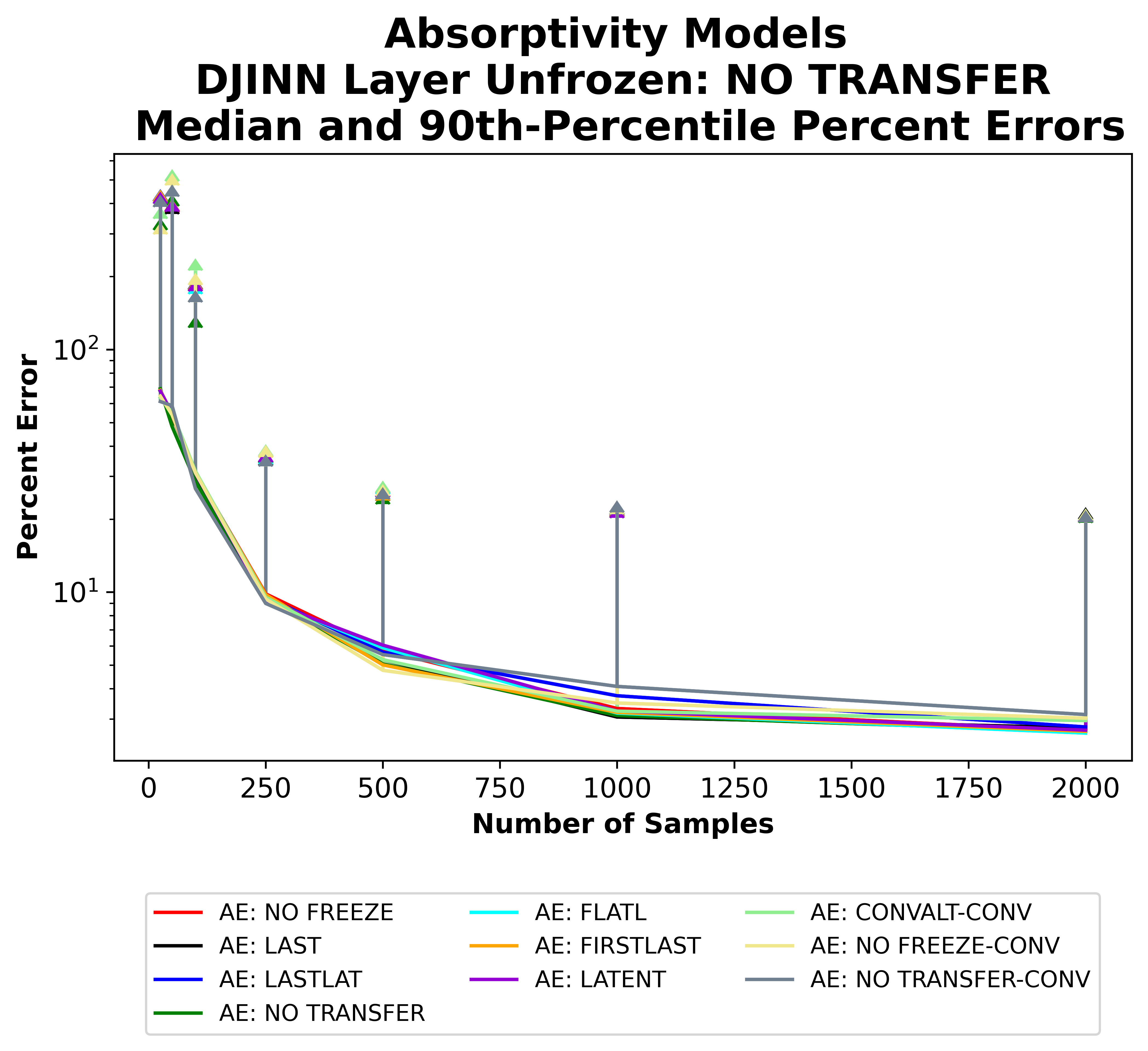}
        \includegraphics[width=\linewidth]{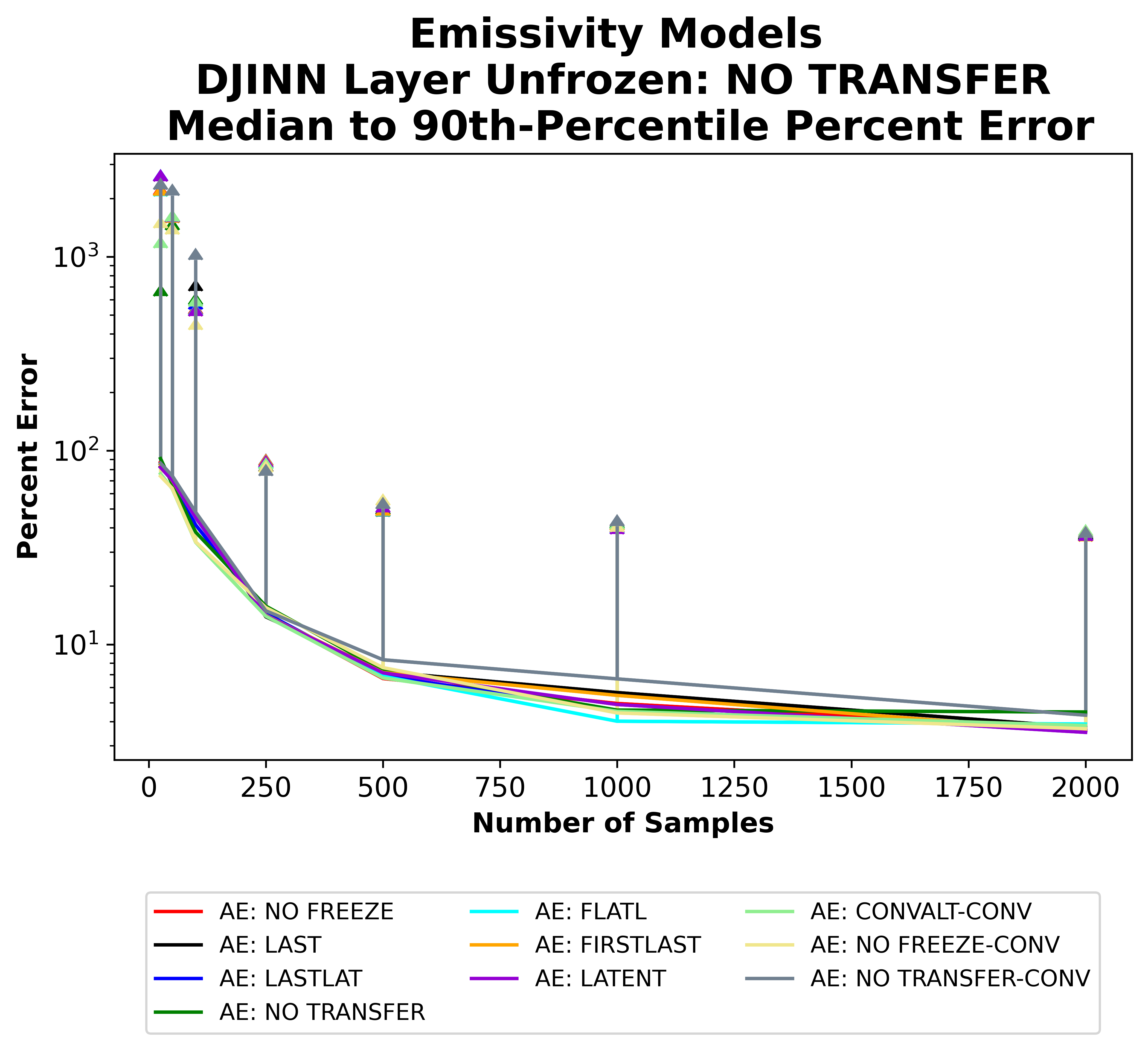}
    \caption{These plots are the results of the models that utilized a brand new, i.e. ``NO TRANSFER'', model. This result provides a demonstration that it is the DJINN model that is primary contributing factor in the error.}
    \label{fig: no transfer}
\end{figure}

The entirely brand new models performed similarly to the models that had a brand new DJINN model made with a transfer learned autoencoder. The results of a brand new DJINN model built from the various autoencoders can be seen in Figure \ref{fig: no transfer}. 
Building the wholly brand new models, i.e., the ``NO FREEZE'' models, requires roughly the same amount of data to achieve the same level of median and 90th-percentile accuracy as the new DJINN models built with transfer learned autoencoders; Figure \ref{fig: no transfer} shows that all of the models follow the same basic curve. That being said, maximum error performance does generally get better with increased available data, and the brand new models do not perform nearly as well in maximum error than the transfer-learned models. The convolutional models still tend to perform better than fully-connected networks for absorptivity, but this is not always true for emissivity.

\begin{figure}[!t]
    \centering
        \includegraphics[width=\linewidth]{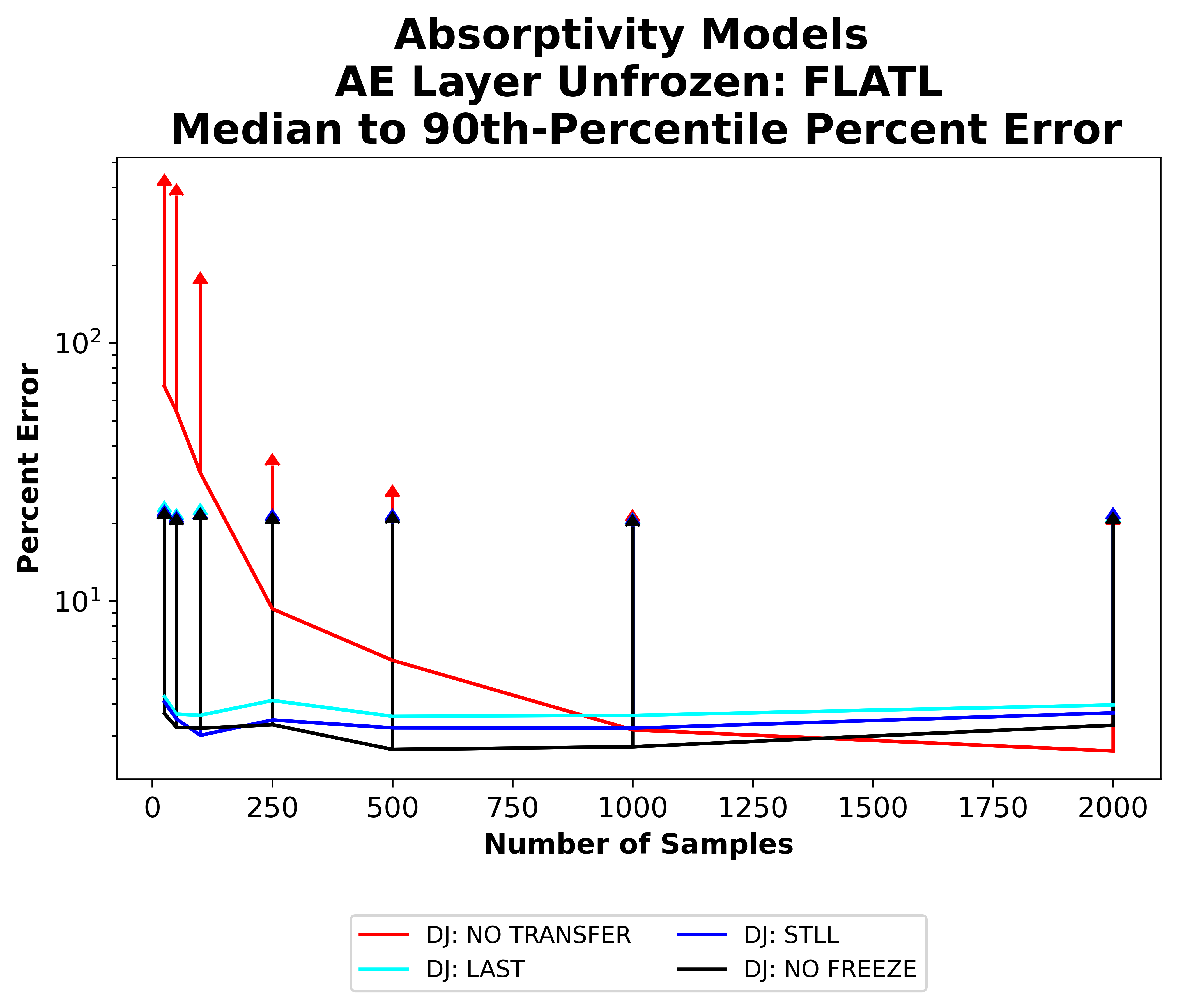}
        \includegraphics[width=\linewidth]{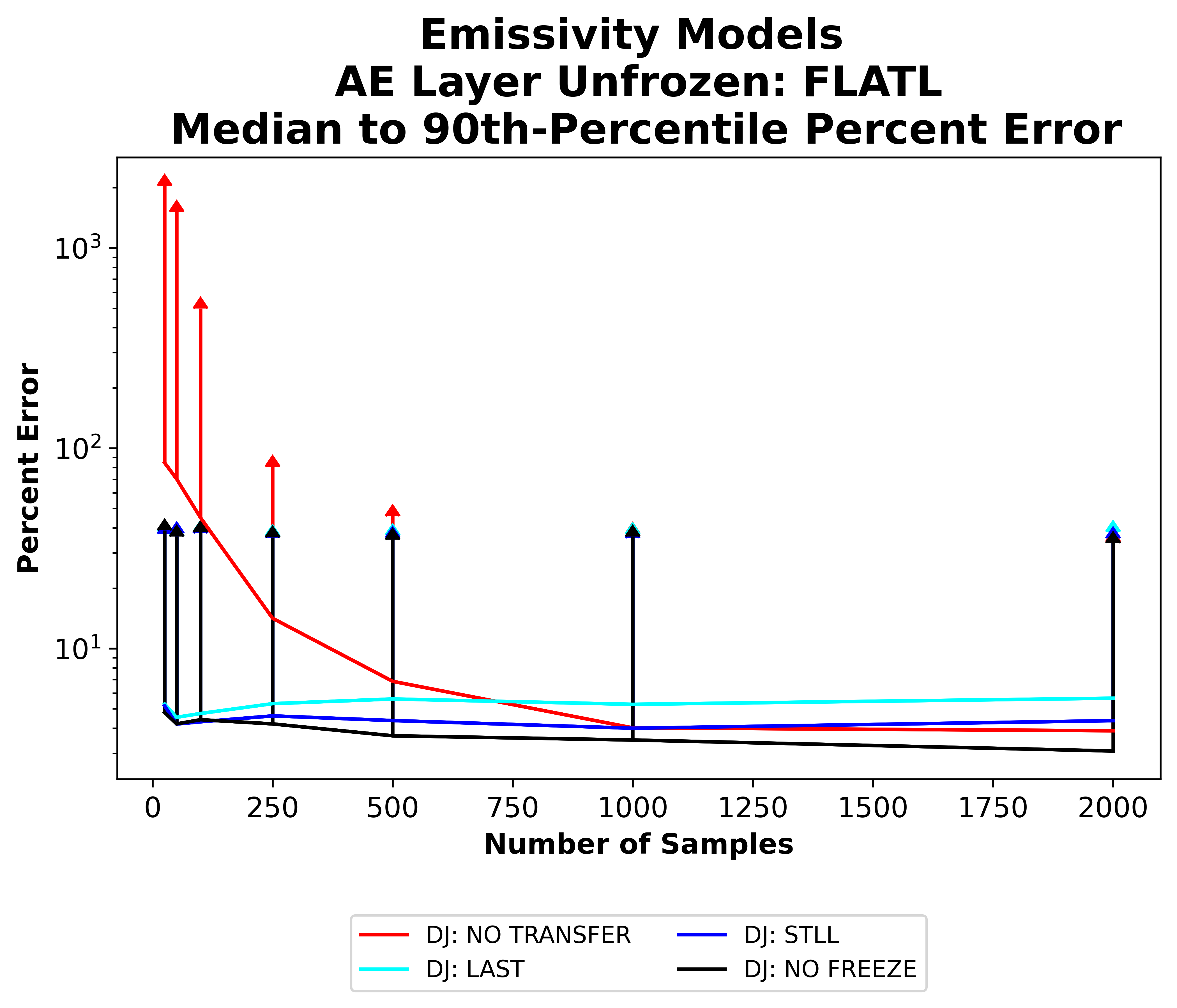}
    \caption{These are the results of the models made with a ``FLATL" autoencoder. The general progression of results are the same for both absorptivity and emissivity where the ``NO TRANSFER" DJINN models were worse than the other models until about 1000 samples were provided, and the other DJINN models produced results that remained relatively flat after about 500 samples were provided. The error values reported are the median values of each reported metric (median or 90th percentile) out of ten separate models.}
    \label{fig:dj freeze compare}
\end{figure}

The results of a brand new DJINN model built from a ``NO FREEZE" autoencoder can be seen in Figure \ref{fig:dj freeze compare} which compares it against the DJINN models built from ``NO FREEZE" autoencoders. It is clear that for both absorptivity and emissivity, the brand new DJINN models do not begin to perform as well as the fully transfer-learned models until about 1000 samples are used. A note to make about the brand new models is that the models made with $<1000$ samples produce predictions that are the correct shape, but those predictions have a large offset from the target data.

These results justify that, based purely on error, transfer learning will be needed if only small amounts of data are available. On a positive note, it does show that only about 1000 samples are necessary to build an acceptably accurate model.

This is more apparent in the absorptivity spectra than the emissivity spectra despite absorptivity having the least amount of error. Most of higher-density ($> 0.01 \text{g/cm}^3$) inputs produce spectra that are only a few percent different between the low-fidelity and high-fidelity data, and the shape is generally very similar for emissivity. The low density inputs produce high-fidelity spectra that are significantly different from the low-fidelity data as seen in Figure \ref{fig:lf compare}. 

\begin{figure}[!t]
    \centering
        \includegraphics[width=\linewidth]{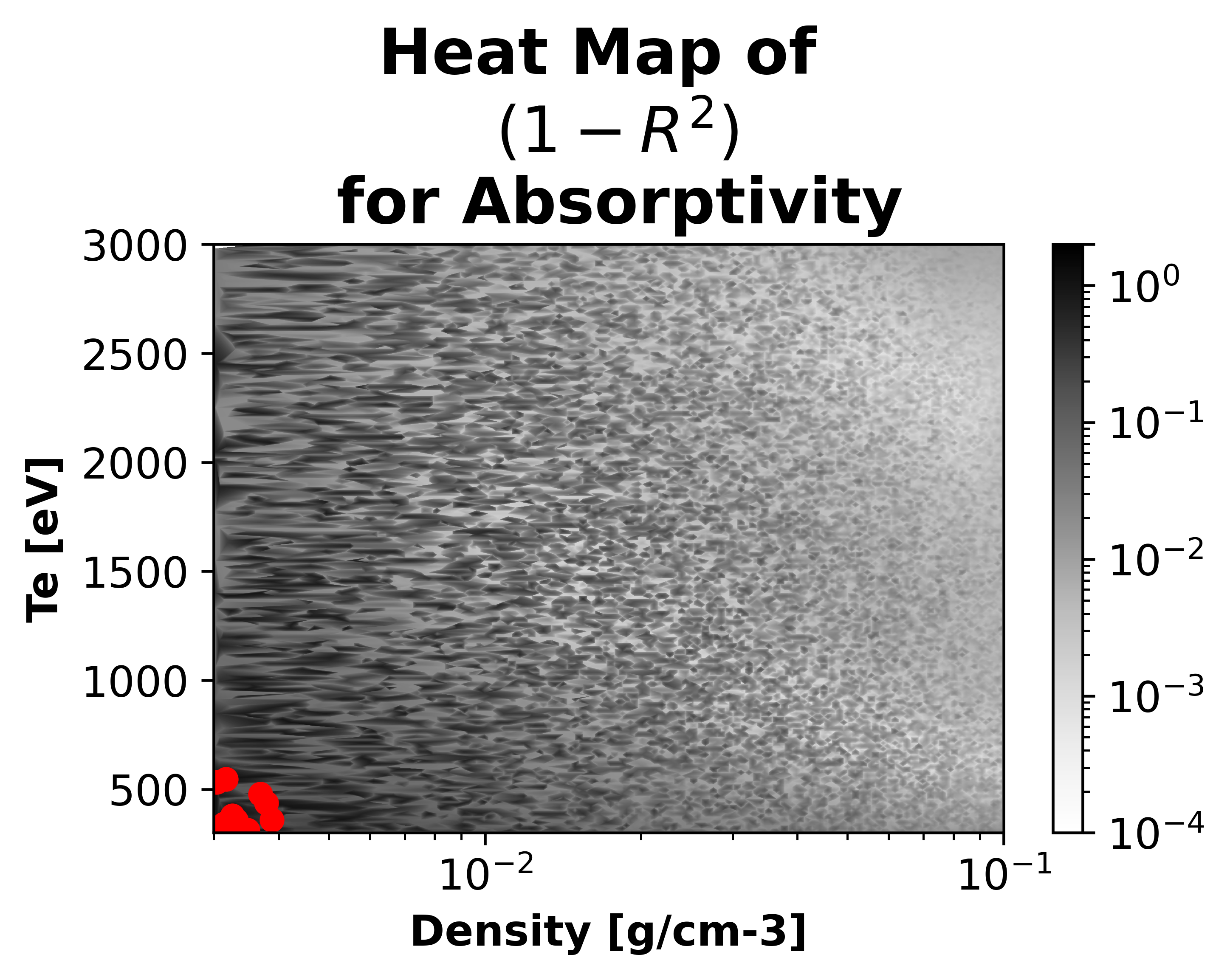}
        \includegraphics[width=\linewidth]{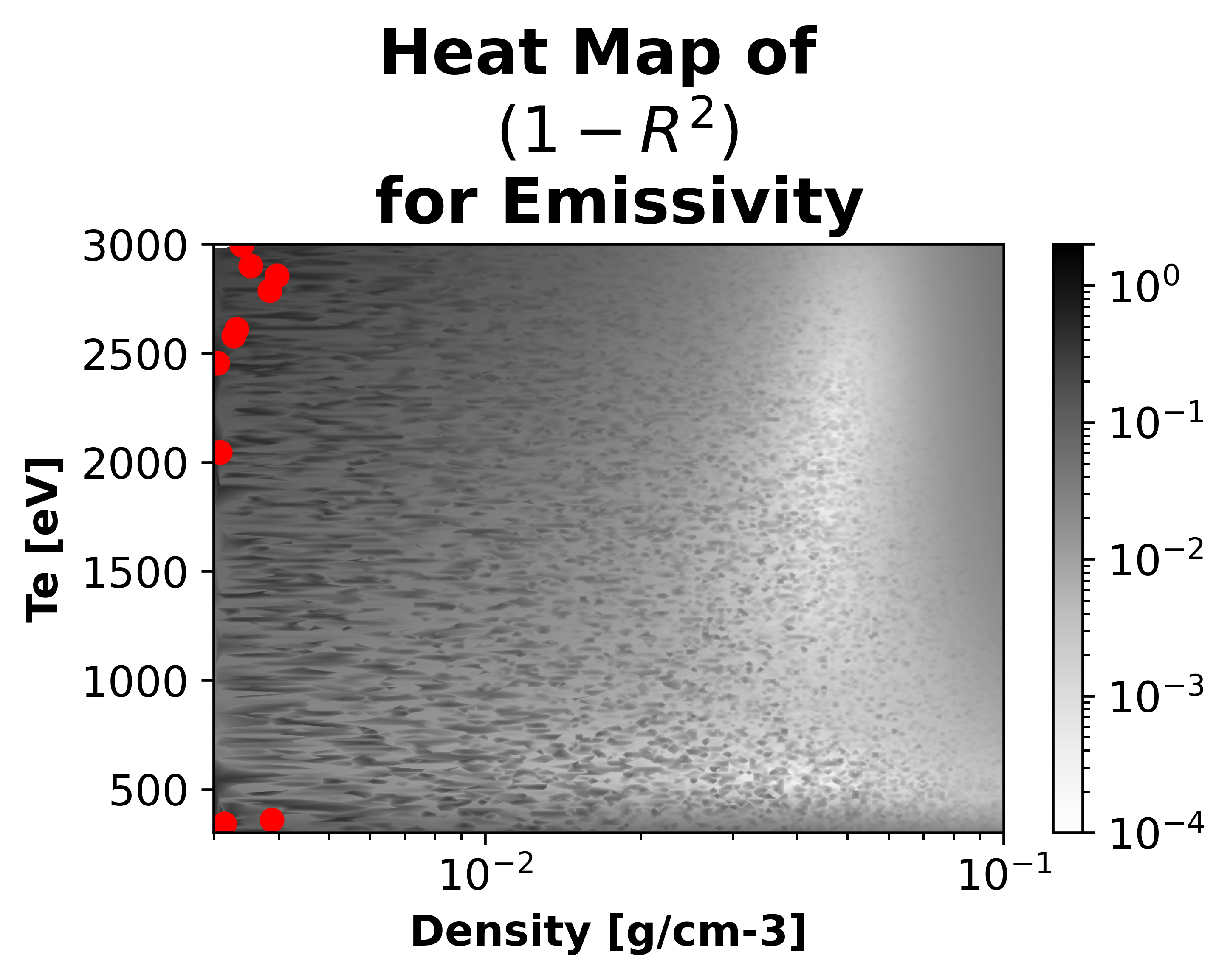}
    \caption{The metric plotted in these heat maps is the mean ($1-\text{R}^2$) for each temperature-density bin across ten models. The ($1-\text{R}^2$) is calculated for each spectra associated with that temperature-density bin. These heat maps are for the models made with a ``NO FREEZE" autoencoder and a ``NO FREEZE" DJINN model. These are from fully-connected models. The red dots are the input locations associated with the ten worst ($1-\text{R}^2$) values.}
    \label{fig:heat maps}
\end{figure}

The error associated with shape is difficult to quantify, but it is inherently embedded in the general numerical error. Thus, a representation of the distribution of errors can be seen in Figure \ref{fig:heat maps}. The heat maps show that both absorptivity and emissivity are primarily dependent on the density with the emissivity appearing to be more so than the absorptivity. This conclusion is based on the less smooth gradient in the heat map for absorptivity. The red dots in the heat maps show that the worst spectral reproductions are concentrated in the lowest temperature ($<500\text{eV}$) and highest temperature ($>2000\text{eV})$ combined with the lowest density ($<.005\text{g/cc}$) regions where density appears to be the most contributing factor.

\changeII{In Figure \ref{fig:heat maps} we notice that the largest errors occur towards the edge of the input ranges for density and temperature. Not depicted in the figure is the other input to the NLTE calculation, the radiative field. The primary reason for the largest errors being at the edges of the training data is a mathematical one: data-driven models (like NNs) typically perform worst near the edges of the training data. There is also a physics reason for the large error samples to exist at the low density portion of the figure. At lower densities, the fewer collisions between atoms cause the atomic to have more detailed absorptivity and emissivity curves. These types of spectra will naturally be harder to reproduce. It is also worth noting that our models give the best results at relatively low density (below 0.1 g/cm$^3$) and at higher temperatures ($> 1$ keV). These conditions should require NLTE physics to describe the radiation interactions. In a radiation hydrodynamics calculation, it may be beneficial to switch to an LTE model when conditions justify that approximation, this would also require considering the local radiative field conditions.  Our results do not indicate that this is necessary to do as the ML models have a behavior that is more stable as the density is increased and can handle variations in the radiative field.}

The smooth line of the low-fidelity data is effectively a line of best fit between all of the peaks and troughs of the high-fidelity data. This probably produces a ``deep" local minima in the model that training cannot be escaped under the same conditions that places the model in that minima.

\begin{figure}
    \centering
        \includegraphics[width=\linewidth]{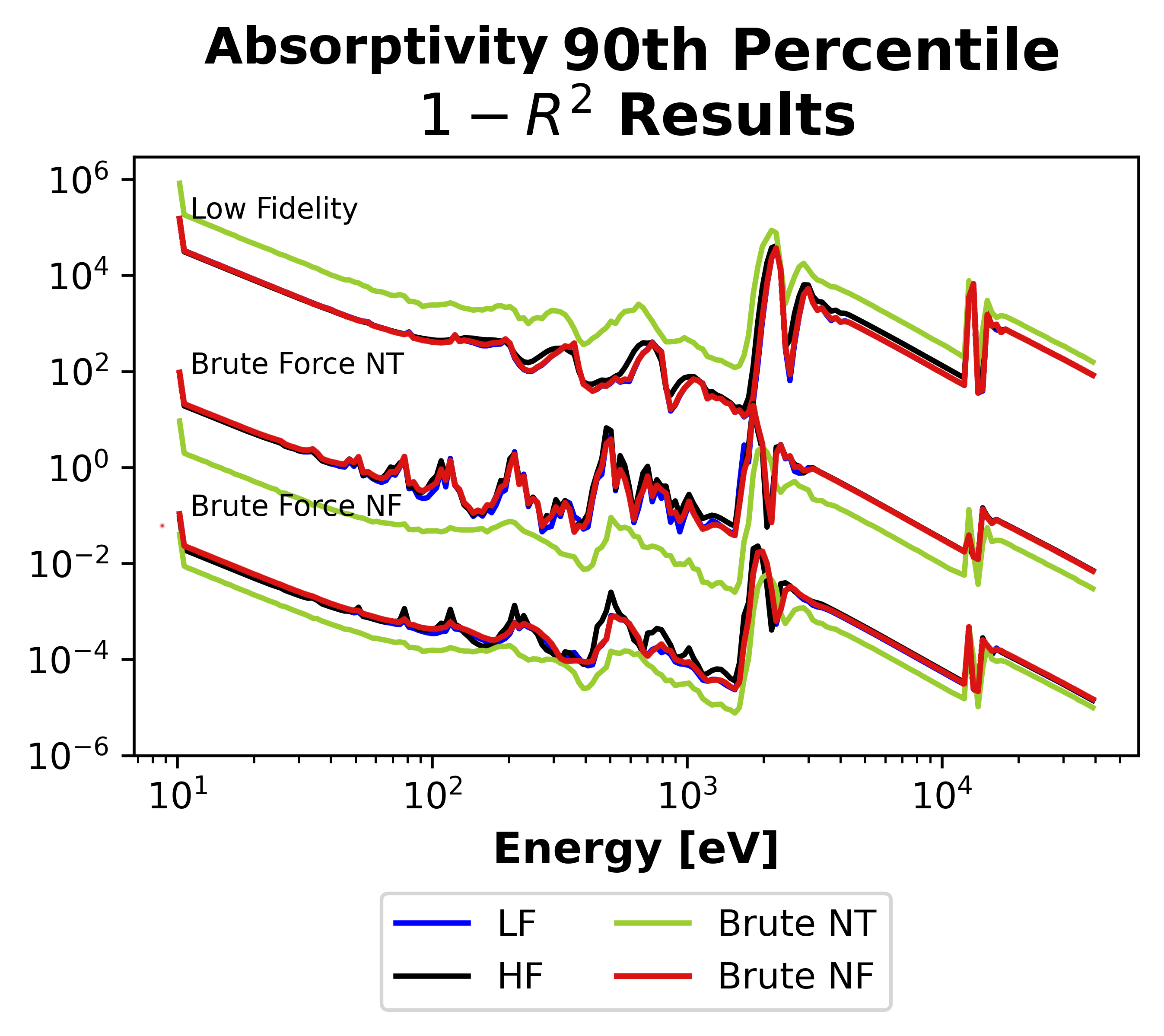}
        \includegraphics[width=\linewidth]{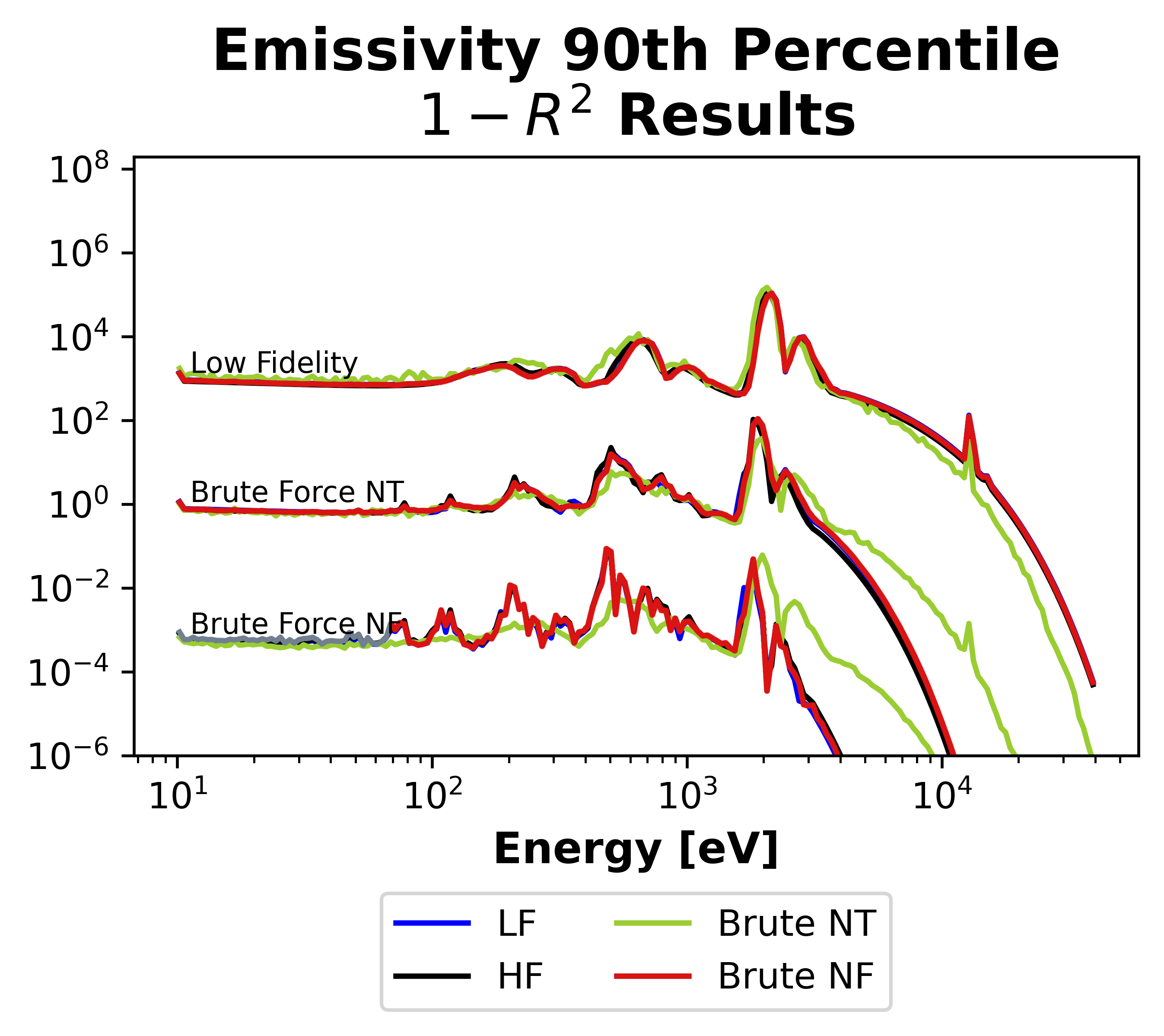}
    \caption{These are the spectra corresponding to 90th-percentile ($1-\text{R}^2$) values of various models. The labels on in the plot refer to which model the 90th-percentile input was obtained from. The 90th-percentile of one model may not be the same as the 90th-percentile of another model. The lines associated with the ``Low-Fidelity" label are based on the inputs that produce the largest difference between the actual low-fidelity and high-fidelity data. ``Brute NT" and ``Brute NF" refers to the ``NO TRANSFER" and ``NO FREEZE" models accordingly. Each set of lines was shifted to improve readability. }
    \label{fig: 90th percentile}
\end{figure}

As with most machine learning tasks, this approach works well, but it is not perfect. The spectra that demonstrate the greatest difference between low-fidelity and high-fidelity prove to be the more difficult spectra to reproduce. However, the bulk majority of spectra are reproduced well. Figure \ref{fig: 90th percentile} shows the outcome of the 90th percentile of the ($1-\text{R}^2$) value. These plots also demonstrate the true effectiveness of transfer learning in that the ``Brute NT" (brute force ``NO TRANSFER") model has a relatively constant offset for absorptivity. The ``Brute NF" (``NO FREEZE") model does not possess this offset. The ``Brute NT" model performs better for emissivity, but it is still outdone by the transfer learned model.

\begin{figure}[!t]
    \centering
        \includegraphics[width=\linewidth]{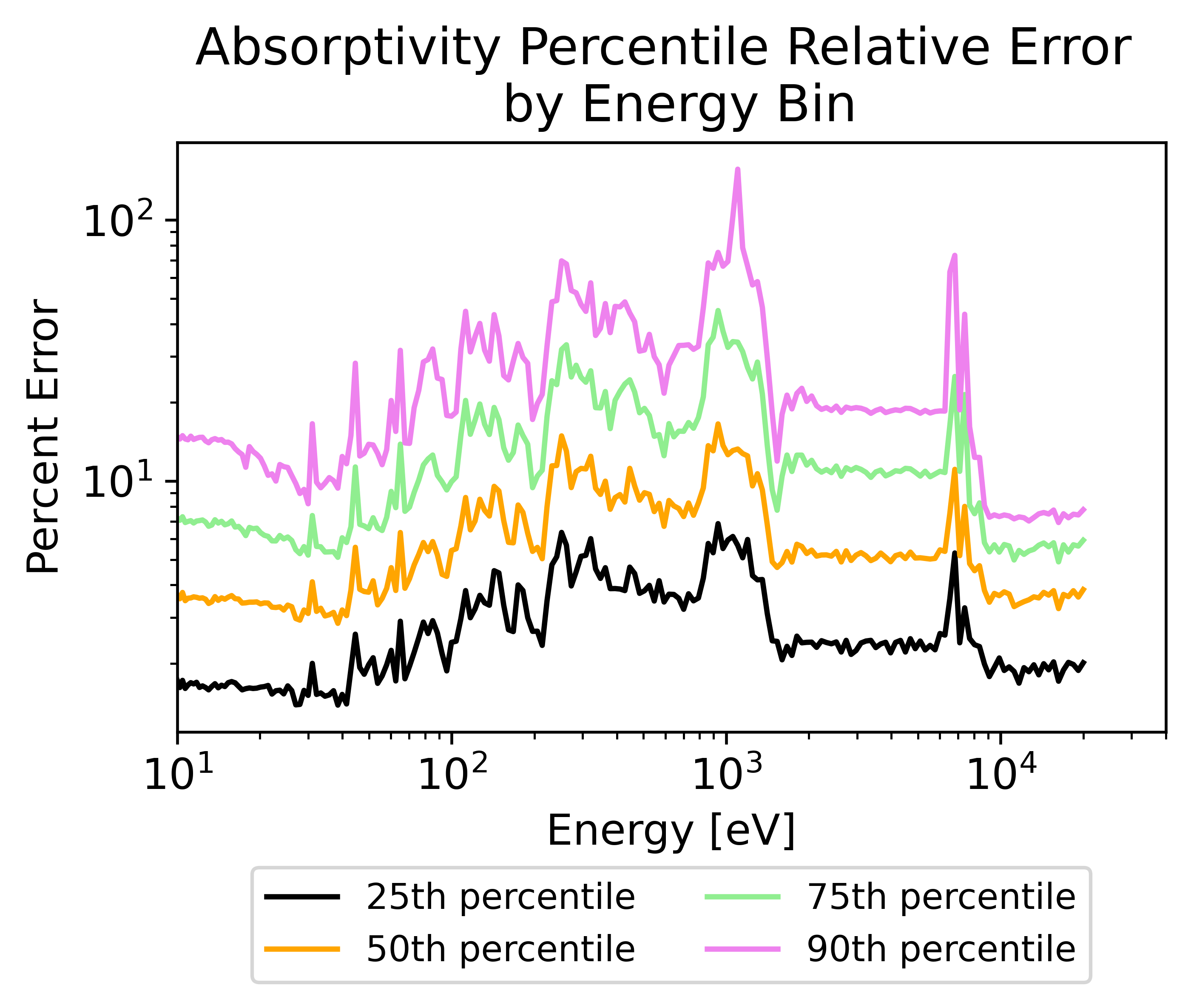}
        \includegraphics[width=\linewidth]{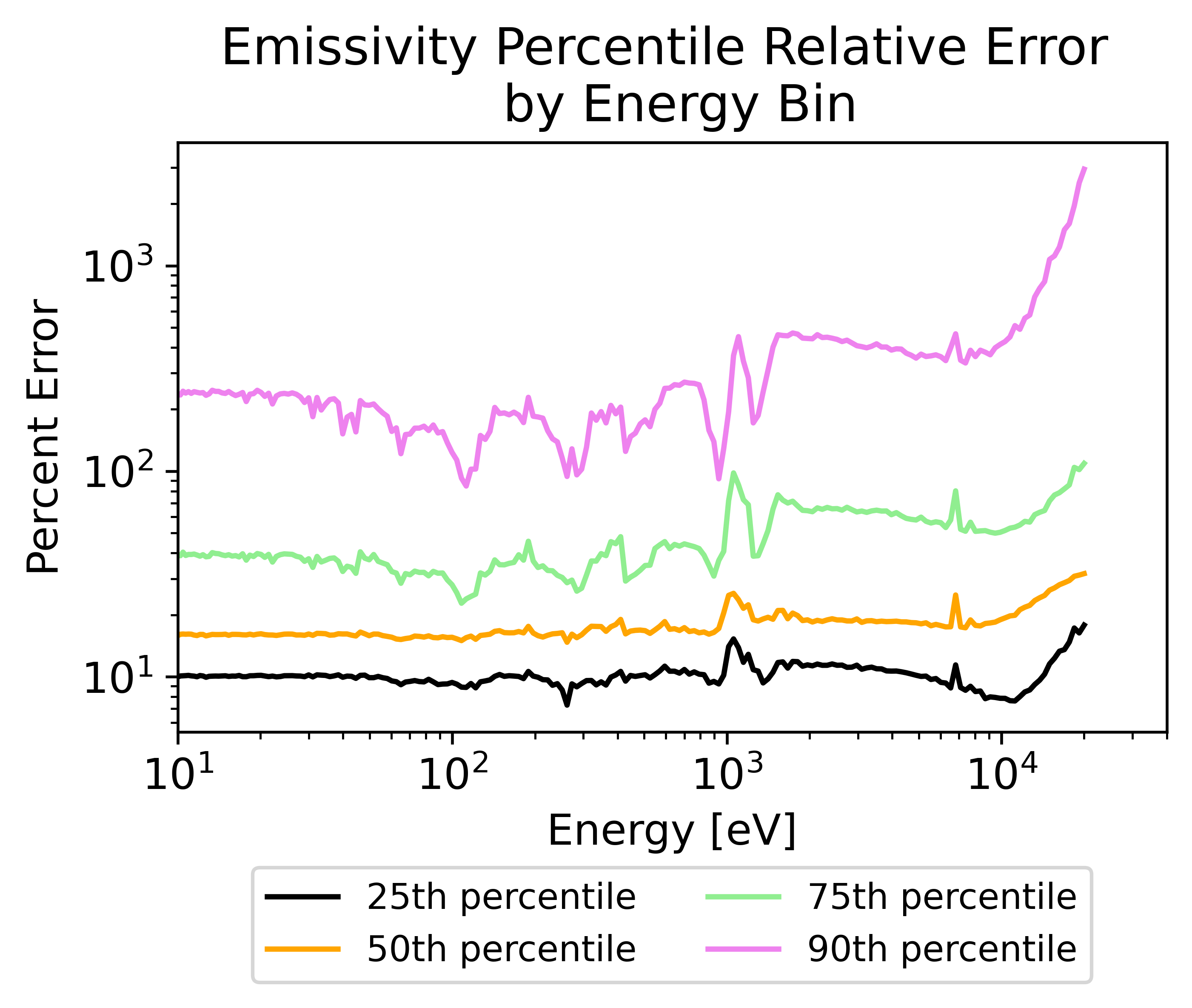}
    \caption{These plots show the 25th, 50th, 75th, and 90th-percentile relative errors for each energy bin the in the spectra for both absorptivity and emissivity. The peaks in relative error correspond to the peaks and troughs in the spectra. The error peaks do not necessarily come from a simple increase or decrease in desired value as some spectral peaks shift in location. }
    \label{fig:spectral error quantiles}
\end{figure}

The errors associated with energy bins where the largest differences in features can also be seen in the overall bin-wise errors. Figure \ref{fig:spectral error quantiles} provides a visualization of where the errors are concentrated among the energy bins. By comparing the plots in Figure \ref{fig: 90th percentile} with the plots in Figure \ref{fig:spectral error quantiles} one can see the regions of the poor shape match appear to be associated with jumps in error which also happens to be where either the spectral peaks or spectral troughs are located. For many of the spectra, the biggest difference between the low-fidelity and high-fidelity is the height of the peaks, which isn't much, but there are certain situations where the the peak actually shifts. The peak in error located around 8 keV-9 keV is the result of one those shifts. This particular case was that the spectral peak need to shift to lower energy bin.

\begin{figure}[!t]
    \centering
    \includegraphics[width=\linewidth]{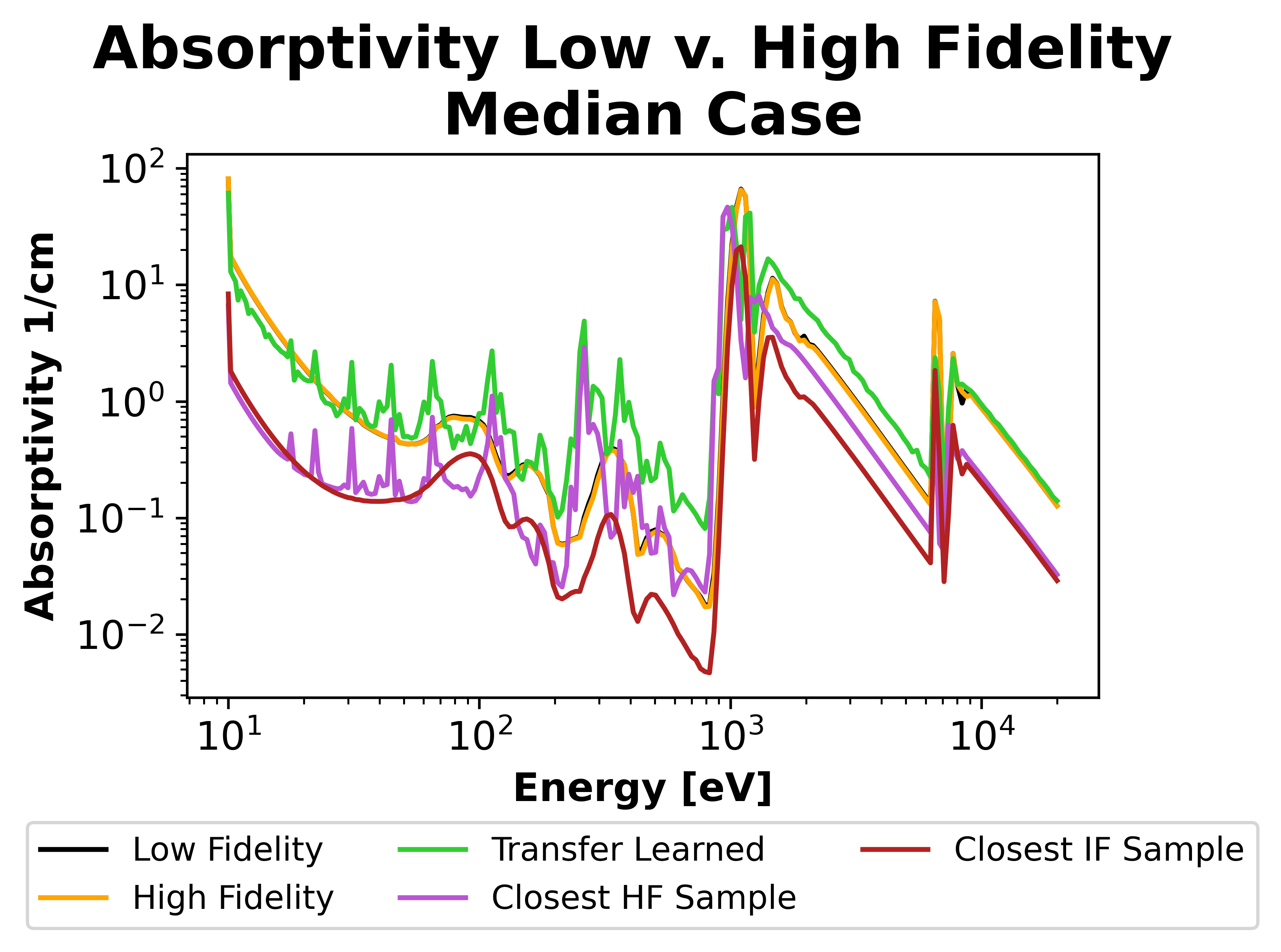}
    \caption{This is a comparison of high-fidelity spectra and its associated low-fidelity spectra against the closest high-fidelity training sample and intermediate-fidelity training sample in terms of the input space. The closest intermediate-fidelity training sample shares the same shape as the desired high-fidelity spectra, but the closest high-fidelity training sample looks more like the low-fidelity counterpart of the desired high-fidelity spectra.}
    \label{fig:diff spectra}
\end{figure}

During the exploration of various methods not mentioned, other information was gleaned during additional attempts to add an intermediate step between the low and high fidelity data it was found that the intermediate fidelity data, which normally takes about a quarter of the time to compute, is almost identical to the high-fidelity data. They are so similar that the lines can hardly be distinguished from each other on a plot. Further, the set of inputs in the training set that are closest in euclidean distance to the inputs corresponding to the actual spectra of the worst prediction, based on the median relative error of the spectra as opposed to the \rsquare{} value, may correspond to a spectra with a shape that is significantly different from the desired shape. Such an example of this can be seen in Figure \ref{fig:diff spectra} where the closest intermediate-fidelity absorptivity spectra matches the shape of the expected absorptivity spectra closely, but the closest high-fidelity spectra is significantly different from the expected spectra. In the exact opposite manner, the closest intermediate-fidelity emissivity is significantly different from from the expected emissivity spectra, but the closest high-fidelity spectra does look similar. The difference in the case of the emissivity model here is that the model actually did a comparatively decent job at matching the spectra's shape. This leads the authors to conclude that there are not enough data points or that random sampling with so few samples is not effective.

\section{Conclusions}

In this paper it has been demonstrated that transfer learning high-fidelity data from a low-fidelity model using only 50 high-fidelity samples can achieve median relative errors in the realm 3\%-4\%. The improvement over using a model trained on only low-fidelity data is about 2x. Also, simply having more samples does not guarantee significantly better error values. It also happens that the barrier to achieving even better error values may very well be the difficulty in matching the shape of the spectra. The following suggestions are based on the authors' experience as well as the results reported in this paper.

The authors' suggest that, as should be expected, the highest level of fidelity that can be reasonably produced in quantities exceeding 10,000 samples should be used as the initial low-fidelity data. The reason for this is that the low-fidelity data here was so smooth that it was most likely the biggest obstacle to achieving better spectral shape reproduction than was seen. Next, if numerical approximation is all that is desired, 50 samples of high-fidelity data appears to be all that is necessary to transfer learn the low-fidelity models to reproduce high-fidelity data.

In terms of the method to be used, a low-fidelity model should first be produced from 10,000 or more samples with temperature being sampled from a uniform distribution and the density being sampled from the uniform distribution on the $log_10$ transform of the density. A fully-connected network should be used for emissivity, and a convolutional model should be used for absorptivity. If using it a convolutional model for emissivity, it should be noted that it will almost certainly produce maximum errors significantly higher than a fully-connected network. However, this error will be located in the high energy region of the spectra where values are extremely low and in a region that may not be of particular concern. In either case, the selection of which layers are frozen does not appear to matter that much, thus for the sake of simplifying implementation, it is suggested that all layers be kept trainable.

Perhaps, just as important as the suggestions provided above is an intimate knowledge of the direct computation models. As discovered with the intermediate-fidelity data briefly mentioned in regards to \ref{fig:diff spectra}, there is almost no difference between it and the high-fidelity data, yet the high-fidelity data requires 4x the amount of time to compute. The difference in the calculated values is well within the expected range of error that might be seen in the approximation by neural network models; however, this may not hold for all binning structures. It is suggested that, in regions of the input space where there is very little difference in the spectra, lower fidelity data be used to directly augment the training dataset effectively making the training data much cheaper to produce. The point at which this should be avoided can only be determined by the user's desired level of error.

Lastly, to improve upon this work, an exploration of whether the premise of only needing 50 samples will hold for different binning structures and resolutions. Further, because this exploration of transfer learning was only done for a \cretin{} surrogate model, this work will have it's implementation in actual \hydra{} simulations tested. Hydra models use actual radiative fields from simulations which need to be compressed with an autoencoder. Finally, it must be determined whether there is any discernible difference in the outcomes of the statsimulations.

\section{Acknowledgements}
This work was performed under the auspices of the U.S. Department of Energy by Lawrence Livermore National Laboratory under Contract DE-AC52-07NA27344 through the  Academic Collaboration Team University Program (ACT-UP). Released as LLNL-JRNL-832026-DRAFT.

This document was prepared as an account of work sponsored by an agency of the United States government. Neither the United States government nor Lawrence Livermore National Security, LLC, nor any of their employees makes any warranty, expressed or implied, or assumes any legal liability or responsibility for the accuracy, completeness, or usefulness of any information, apparatus, product, or process disclosed, or represents that its use would not infringe privately owned rights. Reference herein to any specific commercial product, process, or service by trade name, trademark, manufacturer, or otherwise does not necessarily constitute or imply its endorsement, recommendation, or favoring by the United States government or Lawrence Livermore National Security, LLC. The views and opinions of authors expressed herein do not necessarily state or reflect those of the United States government or Lawrence Livermore National Security, LLC, and shall not be used for advertising or product endorsement purposes.

\bibliographystyle{IEEEtran}
\bibliography{bibs}

\end{document}